\documentclass{emulateapj} 
\usepackage{apjfonts}
\pdfoutput=1

\begin{document}

\title{Unifying the Zoo of Jet-Driven Stellar Explosions}

\shorttitle{Jet-Driven Stellar Explosions}

\author{Davide Lazzati\altaffilmark{1}, Brian
  J. Morsony\altaffilmark{2}, Christopher
  H. Blackwell\altaffilmark{1,3}, \and Mitchell
  C. Begelman\altaffilmark{4,5}} \shortauthors{Lazzati et al.}

\altaffiltext{1}{Department of Physics, NC State University, 2401
Stinson Drive, Raleigh, NC 27695-8202}

\altaffiltext{2}{Department of Astronomy, University of
Wisconsin-Madison, 2535 Sterling Hall, 475 N. Charter Street, Madison WI
53706-1582}

\altaffiltext{3}{Department of Physics, Optics Building, University of
  Alabama in Huntsville, Huntsville, AL 35899, USA}

\altaffiltext{4}{JILA, University of Colorado, 440 UCB, Boulder, CO
80309-0440}

\altaffiltext{5}{University of Colorado, Department of Astrophysical and
Planetary Sciences, 389 UCB, Boulder, CO 80309-0389}

\begin{abstract} 
  We present a set of numerical simulations of stellar explosions
  induced by relativistic jets emanating from a central engine sitting
  at the center of compact, dying stars. We explore a wide range of
  durations of the central engine activity, two candidate stellar
  progenitors, and two possible values of the total energy release. We
  find that even if the jets are narrowly collimated, their
  interaction with the star unbinds the stellar material, producing a
  stellar explosion. We also find that the outcome of the explosion
  can be very different depending on the duration of the engine
  activity. Only the longest-lasting engines result in successful
  gamma-ray bursts. Engines that power jets only for a short time
  result in relativistic supernova explosions, akin to observed
  engine-driven SNe such as SN2009bb. Engines with intermediate
  durations produce weak gamma-ray bursts, with properties similar to
  nearby bursts such as GRB~980425. Finally, we find that the engines
  with the shortest durations, if they exist in nature, produce
  stellar explosions that lack sizable amounts of relativistic ejecta
  and are therefore dynamically indistinguishable from ordinary
  core-collapse supernov\ae.
\end{abstract}


\section{Introduction}

Massive stars end their life with powerful explosions called
supernov\ae (SNe, Woosley \& Weaver 1986).  Some SNe are brighter than
others, and in extreme cases a gamma-ray burst (GRB) - the brightest
explosion in the present universe - is produced (Hjorth et al. 2003;
Stanek et al. 2003; Woosley \& Bloom 2006).  It is believed that a GRB
is associated with a SN when the progenitor star is compact, its
stellar core has fast rotation, and the collapse of the core creates a
fast spinning compact object that releases bipolar jets of
relativistic matter and energy (MacFadyen \& Woosley 1999; Woosley \&
Heger 2006).  Recent observations (Soderberg et al. 2010; Bietenholz
et al. 2010) and theoretical work (Khokhlov et al. 1999; Burrows et
al. 2007; Wheeler \& Akiyama 2010; Nagakura et al. 2012) have shown
that an engine of the same kind may be present in some more mundane
SNe that differ from their normal counterparts only by the presence of
a fast and bright radio transient.

Long-duration GRBs are powered by light, highly relativistic jets
produced by a still mysterious compact object sitting at the center of
an exploding massive star.  The first hurdle that the newly born jet
has to overcome to become a GRB is the crossing of its progenitor
star.  The light, relativistic plasma has to bore a hole through the
cold, dense stellar material without being excessively polluted by the
star's baryons that would slow it down and prevent the release of the
high-frequency photons that characterize the GRB prompt emission
(Piran 1999, Kaneko et al. 2006).  Numerical simulations showed that
the jet forms a bow shock in the stellar material that advances at
sub-relativistic speed, shedding some of its energy in the process
(MacFadyen \& Woosley 1999; Aloy et al. 2000; Zhang et al. 2003, 2004;
Morsony et al. 2007; Mizuta \& Aloy 2009).  The ``wasted'' energy is
accumulated in a hot cocoon that drives shock waves into the star
(Ramirez-Ruiz et al. 2002; Lazzati \& Begelman 2005; Bromberg et
al. 2011b), unbinding it and possibly producing the supernova
associated with the burst (Hjorth et al. 2003; Stanek et al. 2003).
The time it takes a typical GRB jet to cross the star and break out on
the surface is approximately ten seconds (Morsony et al. 2007).
However, the duration distribution of GRBs includes bursts as short as
two seconds, requiring the existence of engines with a duration barely
long enough to power the jet until the breakout time (Kouveliotou et
al. 1983).  Engines with even shorter activity could exist in nature.
However, their detection is challenging since they do not produce a
clear electromagnetic signature such as a GRB.

Observations of weak GRBs such as GRB 980425 associated with SN1998bw
(Galama et al. 1998; Kulkarni et al. 1998), X-ray flares such as
XRF080109 (Mazzali et al. 2008; Soderberg et al. 2008), and
relativistic radio bright supernovae (SN) such as SN2009bb (Soderberg
et al. 2010; Bietenholz et al. 2010) call for diversity in the
properties of the engines and/or the progenitors that power the
explosion.  In this paper we investigate the role of the duration of
the engine activity on the explosive outcome of compact stars with a
GRB-like central engine. Keeping all the jet and progenitor properties
fairly constant, we vary the engine duration, exploring the
consequences of the non-linearity of the jet propagation velocity for
increasing engine luminosities. The paper is organized as follows. In
Section 2 we present our numerical simulations, in Section 3 we
describe our results, in Section 4 we outline an analytical model to
interpret our results, and in Section 5 we discuss the implications
and limitations of our study.

\section{Numerical Simulations}

\begin{figure}
\plotone{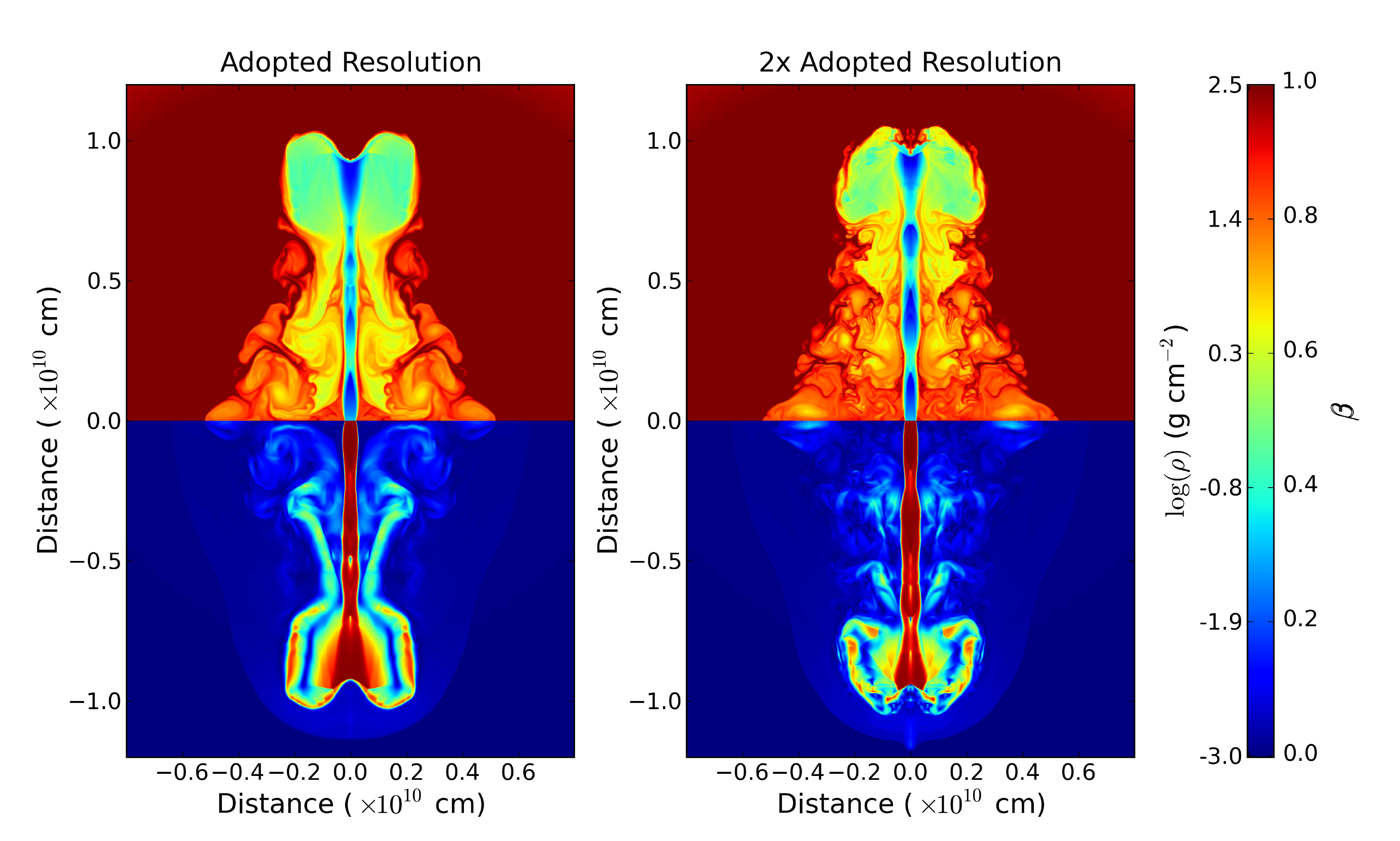}
\caption{{Comparison between jets from identical simulations at the
    adopted resolution (left panel) and at twice the adopted
    resolution (right panel).  The upper part of each panel shows the
    density map in false colors, while the lower part shows the
    velocity map.  The simulations shown have $t_{\rm{eng}}=10.0$~s
    and the frames are taken at $t=6.67$~s after the engine onset.
    The two jets not only have the same size, but also show the same
    features, such as a big turbulent eddies on both sides of the
    head.}
\label{fig:resolution}}
\end{figure}

All the simulations presented in this paper were performed with the
FLASH code (Fryxell et al. 2000).  We adopted a minimum resolution of
$4\times10^6$~cm at the highest level of refinement.  At this
resolution the transverse dimension of the injected jet is resolved
into 44 elements.  This resolution was chosen after verifying that
simulations performed at twice the resolution would yield consistent
results (Figure~\ref{fig:resolution}).  Our simulations do not include
magnetic fields, due to the technical challenge of performing MHD
calculations with relativistic motions on an adaptive mesh.  Our
simulations do not consider nuclear burning since its inclusion would
have made the simulations too long to perform at the adequate
resolution. As a consequence, ejection velocities are marginally
affected.  Finally, gravity from a central mass and self-gravity are
neglected since the characteristic time-scales of the jet-star
interaction are much shorter than the dynamical time of the progenitor
star's collapse.  Due to the absence of gravity, our progenitor star
expands by a modest 2\% of its original size in the 100 seconds of the
simulations. The dynamical time scale at the stellar surface is
$\tau_{\rm{dyn}}\simeq10^5~$s, while at our inner boundary it is
$\tau_{\rm{dyn}}\simeq10$~s, comparable to the break-out time of our
jets.  All our simulations adopted realistic GRB stellar progenitors:
models 16TI and 12OM from Woosley \& Heger (2006).  Model 16TI is a 16
solar-mass Wolf-Rayet star with an initial metallicity 1\% solar and
angular momentum $L=3.3\times10^{52}$~erg~s.  The mass of the star at
pre-explosion is 13.95 solar masses and its radius is
$4.1\times10^{10}$~cm, corresponding to 0.6 solar radii.  Model 12OM
is a 12 solar-mass Wolf-Rayet star with an initial metallicity 10\%
solar and angular momentum $L=2.5\times10^{52}$~erg/s.  The mass of
the star at pre-explosion is 9.5 solar masses and its radius is
$4.8\times10^{10}$~cm, corresponding to 0.7 solar radii.

In all cases, a relativistic jet with opening angle
$\theta_0=10^\circ$ was injected as a boundary condition at a distance
$r_0=10^9$~cm from the stellar center. The engine luminosity varied
among simulations depending on the duration of the engine activity in
order to keep the total energy budget fixed to $E=3\times10^{51}$~erg
(one set of simulations for the 16TI and 12OM progenitor models) or
$E=10^{52}$~erg (one set of simulations for the 16TI progenitor). In
each simulation, the engine luminosity was kept constant until the
cutoff time, when the engine was abruptly turned off. All jets were
injected with Lorentz factor $\Gamma_0=5$ and with enough internal
energy to reach an asymptotical $\Gamma_\infty=400$ upon complete,
non-dissipational acceleration. This jet configuration was used
numerous times to reproduce the properties of successful GRBs (Morsony
et al. 2007, 2010; Lazzati et al. 2009, 2011). A false color still of
the density and expansion velocity of one of our simulations is shown
in Figure~\ref{fig:f1}.

\section{Results}

The simulations were analyzed for the jet dynamics, the presence of an
associated stellar explosion, and the nature of the ensuing explosion.
All our simulations resulted in the explosion of the progenitor star.
However, the fraction of the mass that was ejected into the
interstellar medium varied with the engine activity time.  The shorter
$t_{\rm{eng}}$ was, the higher the fraction of the star mass that
achieved escape velocity.  Analyzing the frames at 15~s after the
engine onset for our simulations of a the 16TI progenitor with
$E=3\times10^{51}$~erg, we found that 95\% of the progenitor mass had
been ejected by our shortest-duration engine ($t_{\rm{eng}}=3.0$~s),
while 88\% of the progenitor mass had been ejected by our longest
engine ($t_{\rm{eng}}=15.0$~s).  We also found that the stellar
explosions are not directly due to the jet, which occupies only a
small fraction of the solid angle, similar to results obtained for
non-relativistic jets (Khokhlov et al. 1999).  The jet propagation,
however, produces a hot, high-pressure cocoon that drives a conical
shock into the stellar material.  For that reason, the star does not
explode spherically.  At the time of the jet breakout, the
star-exploding shock reaches the pole of the star but is still less
than halfway through along the equator.  For example, the equatorial
breakout took three times longer than the polar one in our
$t_{\rm{eng}}=6.0$~s simulation. Computing the fraction of ejected
mass from a simulation that does not include gravity is non-trivial
and, to some extent, not rigorous. At any given time in the simulation
we proceeded as follows to evaluate whether a parcel of matter located
at radius $r$ is bound or not. We first computed the amount of mass
inside a sphere of radius $r$ and evaluated the escape velocity from
such a mass distribution. Subsequently, we compared the escape
velocity with the radial component of the velocity of the parcel of
matter. If the parcel velocity is larger than the escape velocity we
label the matter as unbound, otherwise we consider it still bound,
eventually falling back to the stellar remnant. This procedure has two
limitations. First the mass distribution is not spherically symmetric
and therefore the simple approximation that the gravity on a point
mass only depends on the mass in the inside sphere is not
correct. Second, the lack of gravity in the simulation affects the
mass velocity itself. As a consequence, the values of the unbound mass
fraction reported above should be considered as indicative.

The difference in the explosive outcome is due to the non-linearity in
the jet propagation.  If the propagation velocity of the jet head
would scale linearly with the engine luminosity, explosions with the
same energy budget would look alike since the jet would either stall
inside the star or breakout, independently of the engine luminosity
and duration.  However, we find that the jet head propagates at a
speed that scales less than linearly with the luminosity
(Figure~\ref{fig:f2}, Section~4).  The non-linearity is brought about
by the feedback between the jet and the high-pressure cocoon that
surrounds it.  A more luminous jet has higher internal pressure that
makes its head wider and therefore its propagation harder, shocking a
larger fraction of the stellar material.  This, in turn, makes the
cocoon more energetic and the higher cocoon pressure squeezes the jet
head facilitating its propagation.
 
\begin{figure}
\plotone{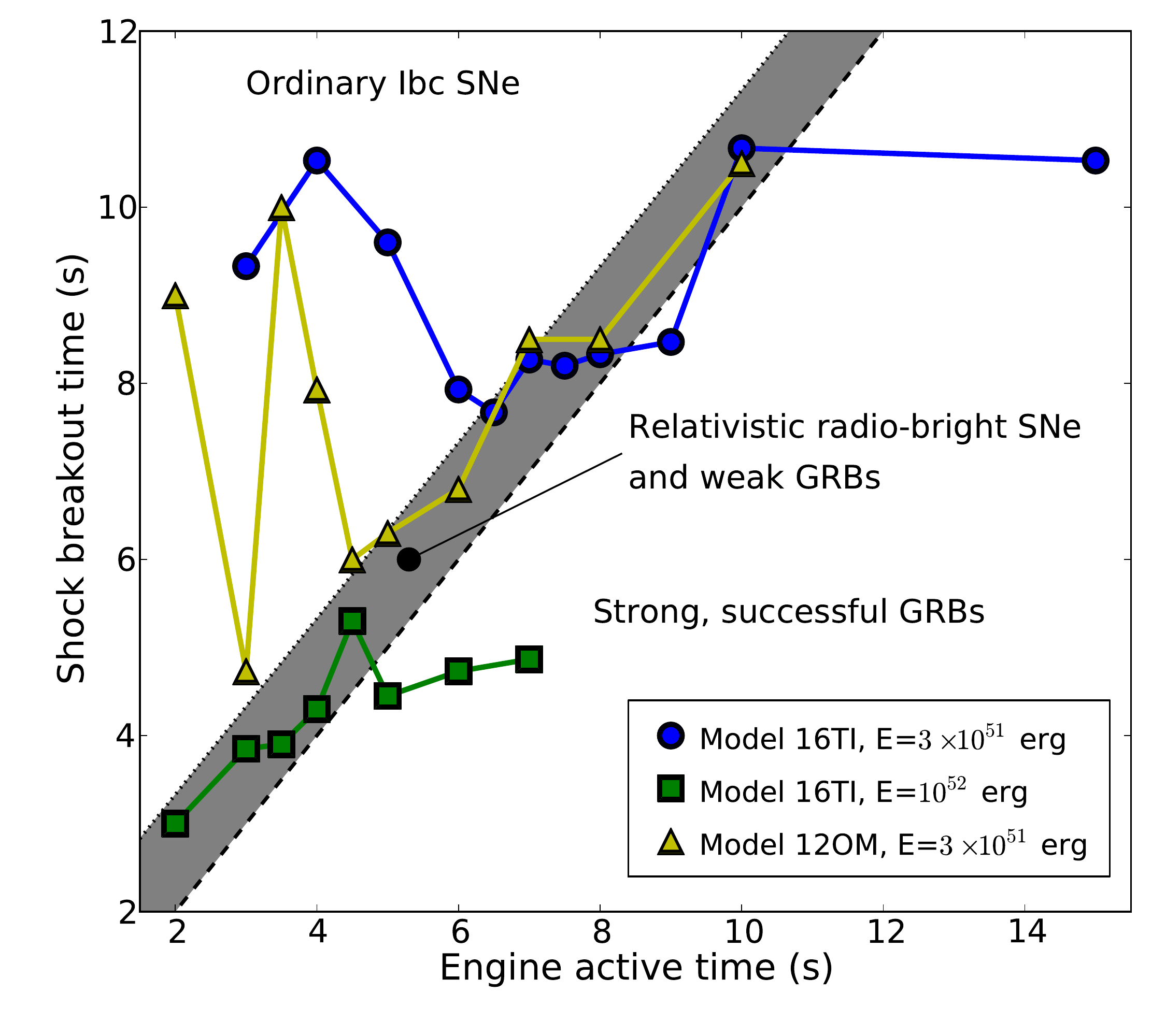}
\caption{{Breakout time of the jet/bubble.  The black dashed line
    shows the condition for which the breakout is at the same moment
    at which the engine turns off.  The black dotted line shows the
    condition for which the breakout takes place after the engine
    turns off but before the tail of the jet catches up with the jet
    head.}
\label{fig:f3}}
\end{figure}

The different jet propagation velocities imply that in some
configurations the jet breaks out the star's surface, while in others
it does not.  We find that there are three conditions that have
relevance to the star's explosion outcome (Figure~\ref{fig:f3}).  If
the engine activity time is short enough, the engine turns off while
the jet head is still buried inside the star.  As the engine turns
off, the tail of the jet detaches from the engine and eventually
catches up with the jet's head.  If this happens while the jet head is
inside the star, all the bulk relativistic motion is lost and the
star's explosion is entirely driven by the high-pressure cocoon that
is left, resulting in an ordinary-looking supernova with little energy
in relativistic ejecta.  This condition is shown with a dotted line in
Figure~\ref{fig:f3}.  Alternatively, the engine may turn off while the
jet head is inside the star but, by the time the tail of the jet has
caught up with the head, the head has already broken free.  In this
case a small fraction of the material moving with bulk relativistic
speed survives and, depending on the details, produces observable
signals in the form of either a bright radio transient or a weak,
GRB980425-like burst.  This condition is shown in Figure~\ref{fig:f3}
with a dashed line.  Finally, if the engine active time is long
enough, the jet breaks out while the engine is still active.  In this
case a fully developed GRB is expected.  Figure~\ref{fig:f3} shows
that all these conditions are realized by an engine with constant
energy but varying activity time.  The comparison between three sets
of simulations shows that the engine duration that characterizes each
type of explosion depends also on the total energy budget and on the
progenitor structure.  The density and velocity maps of a simulation
that produces the intermediate outcome (16 solar-mass progenitor,
$E=3\times10^{51}$~erg, $t_{\rm{eng}}=7.5$~s) are shown in
Figure~\ref{fig:f1}.

\begin{figure}
\plotone{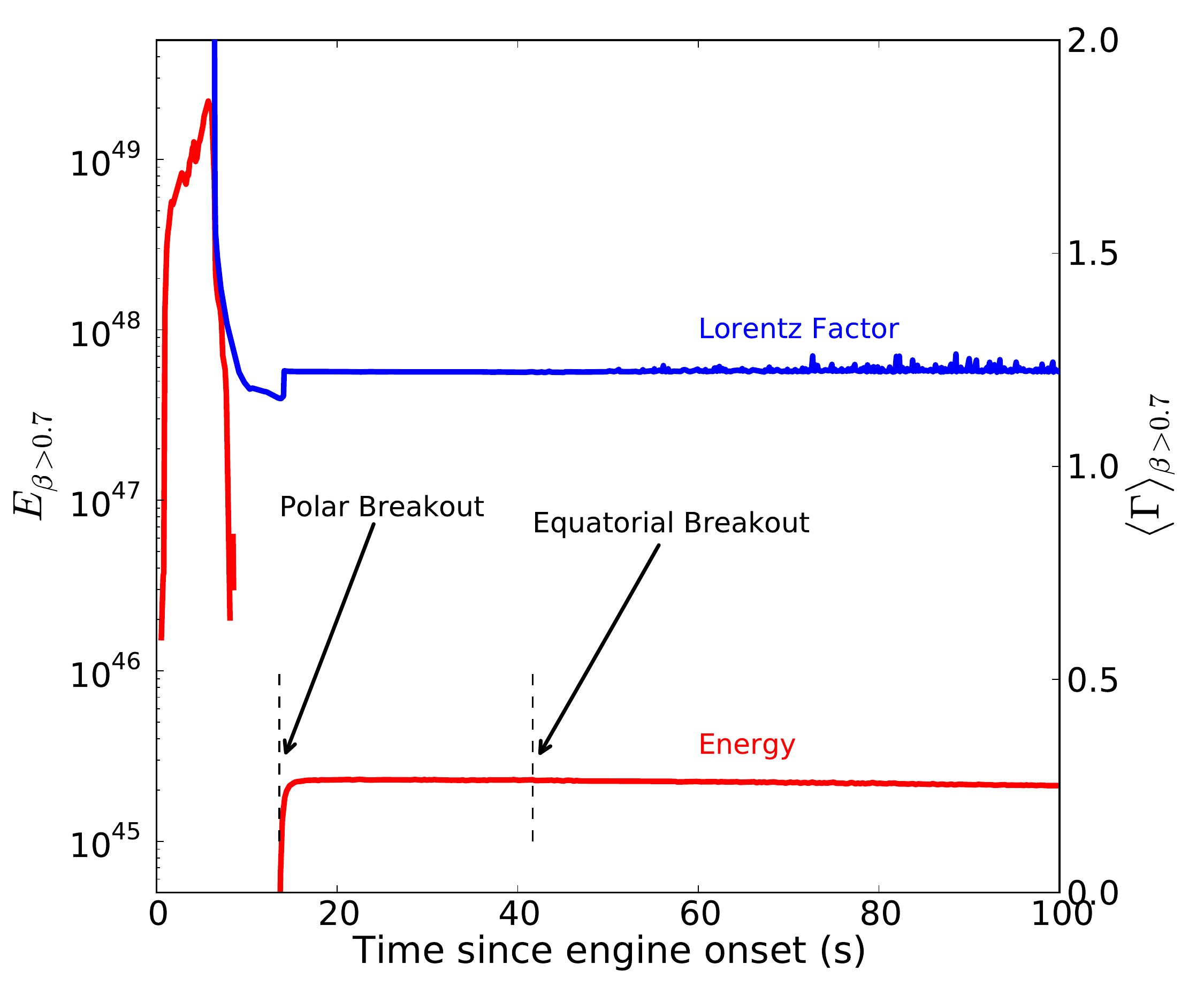}
\caption{{Energy in relativistic ejecta (red) and average Lorentz
    factor (blue) as a function of time after the engine onset for a
    simulation with $t_{\rm{eng}}=6.0$~s and $E=10^{51}$~erg.  The
    figure shows that measuring the energy and Lorentz factor just
    after the shock breakout along the polar axis yield robust
    results, even if the shock breakout along the equatorial plane has
    not yet taken place.}
\label{fig:f6}}
\end{figure}

Figure~\ref{fig:f4} shows a quantitative comparison between the
results of our simulations and the properties of type Ibc SNe and
jet-driven stellar explosions as derived from observations of their
associated radio transient (Soderberg et al. 2010).  Non-thermal radio
emission from SNe is produced by the interaction of the
highest-velocity ejecta with the surrounding ambient medium (Chevalier
1998; Weiler et al. 2002; Chevalier \& Fransson 2006).  We selected
the ejecta moving with $\beta>0.7$ (where $\beta$ is the velocity in
unit of the speed of light) as the material contributing to the radio
transient and we computed the expansion speed as the energy-weighted
average of their velocity. Note that the computation was performed for
$\beta$ even though $\beta\Gamma$ is plotted in
Figure~\ref{fig:f4}. The analysis was performed in the last frame that
we were able to compute, either at $t=100$~s or at the time at which
FLASH would run into a numerical instability (often occurring around
the time of the shock breakout in the equatorial direction). The
energy in relativistic ejecta is a fairly robust number, as long as it
is computed after the jet/cocoon breakout along the polar
direction. Figure~\ref{fig:f6} shows the evolution of the energy in
the ejecta of one of our longest runs as a function of time after the
engine onset. The time of the jet/cocoon breakout along the polar and
equatorial directions are shown. We also checked that the value of the
energy in relativistic ejecta has a very small dependence on whether
the actual expansion velocity or the asymptotic one are considered
(the asymptotic expansion velocity is the one attained once all the
internal energy has been converted in bulk outward velocity). In all
cases the energy once the asymptotic velocity is attained is larger by
less than 10\% with respect to the one shown in the figures. This is
consistent with the fact that the energy in Figure~\ref{fig:f6} is
constant. Should significant acceleration take place, a marked
evolution in the energy would be observed.  In
Figure~\ref{fig:edistri} we show instead the entire cumulative
distribution of the energy of the ejecta for three representative
cases. All the simulations shown are performed with the 16TI
progenitor and have total energy $E=3\times10^{51}$~erg.

\section{Jet propagation model}

The propagation speed of the head of the jet can be found by enforcing
pressure balance along the discontinuity between the jet head and the
progenitor star material.  The pressure balance reads (Matzner 2003;
Bromberg et al. 2011b):
\begin{equation}
\rho_j h_j \Gamma_j^2 \Gamma_h^2 (\beta_j-\beta_h)^2+P_j =
\rho_a h_a \Gamma_h^2 \beta_h^2+P_a
\label{eq:peq}
\end{equation}
where $\beta$, $\Gamma$, $\rho$, $P$, and $h=1+4p/\rho{}c^2$ are the
velocity in units of the speed of light, the Lorentz factor, the mass
density, the pressure, and the dimensionless specific enthalpy,
respectively. An adiabatic index $\hat{\gamma}=4/3$ is used in the
enthalpy expression.  The subscript $j$ refers to the jet material,
the subscript $h$ refers to the jet head, and the subscript $a$ refers
to the ambient medium, in our case the progenitor star.  The above
equation is greatly simplified in the case of a hot, light
relativistic jet slowly advancing into a cold, high-density ambient
medium.  In that case, the pressure terms can be neglected and we can
use the approximations $\Gamma_h=1$, $\beta_h\ll\beta_j$, $\beta_j=1$,
and $P_j\gg\rho_jc^2$.  These simplifications yield: $\rho_j h_j
\Gamma_j^2 = \rho_a h_a \beta_h^2$ which, solving for $\beta_h$ and
using the definition of $h$ becomes:
\begin{equation}
\beta_h = \sqrt{\frac{4P_j}{\rho_a c^2}} \Gamma_j
\label{eq:bh}
\end{equation}

Under the assumption that the pressure of the jet at the head is in
equilibrium with the pressure of the cocoon that surrounds the jet, we
have $P_j=P_c=E_c/3V_c$ where $P_c$, $E_c$, and $V_c$ are the cocoon
pressure, energy, and volume, respectively.  Under our assumption of
sub-relativistic speed of the jet head, the cocoon energy is given by
the energy ejected by the central engine: $E_c=L_jt=L_jr_h/c\beta_h$,
where $L_j$ is the engine luminosity and $r_h$ is the distance
travelled by the jet head.  The cocoon volume is given, assuming an
ellipsoidal shape, by:
\begin{equation}
V_c = \frac{4\pi}{3} r_h r_\perp^2 = \frac{4\pi}{3} r_h (v_\perp t)^2
= \frac{4\pi}{3} \frac{P_c}{\rho_a} \frac{r_h^3}{c^2\beta_h^2}
\label{eq:vcoc}
\end{equation}
where $r_\perp$ is the transverse size of the cocoon, $v_\perp$ is the
velocity of the shock driven by the cocoon into the stellar material,
and we have used the Kompaneets approximation in the last step
(Begelman \& Cioffi 1989; Lazzati \& Begelman 2005).  Using
Eq.~\ref{eq:vcoc} we find the cocoon pressure as:
\begin{equation}
P_c = \sqrt{\frac{cL_ j \rho_a \beta_h}{4\pi r_h^2}}
\label{eq:pcoc}
\end{equation}

\begin{figure}
\plotone{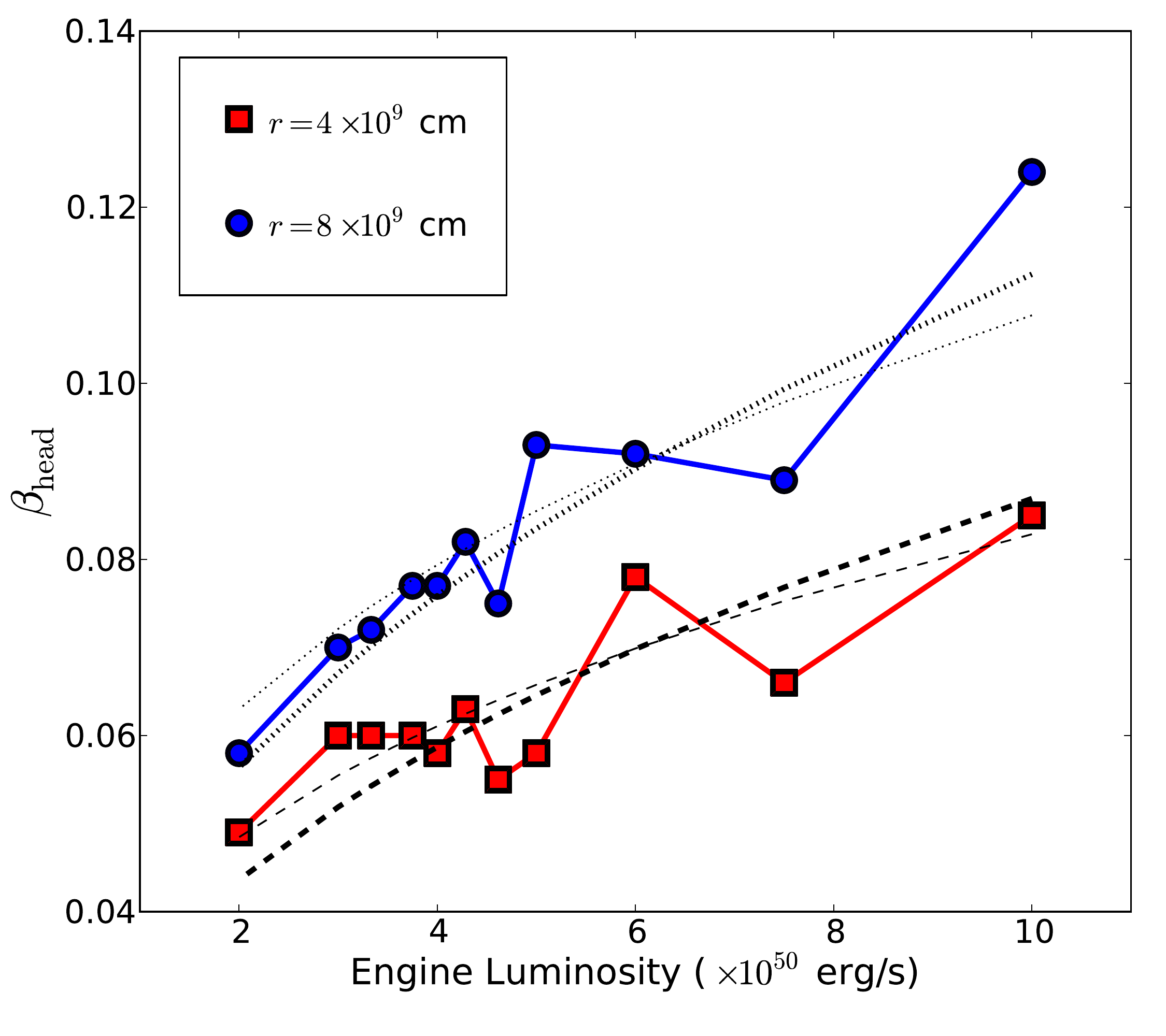}
\caption{{Velocity of propagation of the jet head in the stellar
    material at different stages of the jet-driven explosion.  The red
    curve with squares show the propagation close to the core of the
    star, as the jet head crosses the distance $r=4\times10^9$~cm.
    The blue curve with dots show the velocity as the jet head crosses
    the distance $r=8\times10^9$~cm.  Both datasets are obtained from
    the 16 solar-mass progenitor simulation with total energy
    $E=3\times10^{51}$~erg. A thick dashed line and a thick dotted line with
    equation $v_h\propto{}L_j^{3/7}$ are overlaid on the velocity data
    for comparison with the analytical prediction of
    Eq.~\ref{eq:final}. Thin dashed and dotted lines show instead the
    comparison with the prediction of Bromberg et al. (2011).}
\label{fig:f2}}
\end{figure}

Finally, we need to find a relation between the jet pressure and
Lorentz factor.  The jet pressure is $P_j=L_j/(4c\Sigma_j\Gamma_j^2)$,
where $\Sigma_j$ is the transverse cross section of the jet.  In case
of non-dissipational acceleration, the jet Lorentz factor scales
linearly with its transverse size, and we obtain:
\begin{equation}
\Gamma_j = \left(\frac{L_j\Gamma_0^2}{4c\Sigma_0P_j}\right)^{1/4}
\label{eq:gamma}
\end{equation}
where $\Sigma_0$ and $\Gamma_0$ are the initial transverse cross
section and Lorentz factor of the jet, respectively.  The system of
equations \ref{eq:bh}, \ref{eq:pcoc}, and \ref{eq:gamma} can now be
solved for the jet head propagation velocity:
\begin{equation}
\beta_h = \left(\frac{4\Gamma_0^4}{\pi^3c^9\rho_a^3r_0^4
\theta_0^4}\right)^{1/7}
\frac{L_j^{3/7}}{r_h^{2/7}} 
\label{eq:final}
\end{equation}
where we have used the identity $\Sigma_0=\pi r_0^2\theta_0^2$, $r_0$
being the jet injection radius and $\theta_0$ its initial opening
angle.  The head velocity from Eq.~\ref{eq:final} is compared to the
simulation with thick black lines (dashed and dotted) results in
Figure~\ref{fig:f2}.  We find that Eq.~\ref{eq:final} reproduces
fairly well the dependence of the head velocity on the jet luminosity.
However, it overpredicts the head velocity by a factor $\sim4$.  This
discrepancy is likely due to the approximation made in
Eq.~\ref{eq:gamma}, where it was assumed that the jet accelerates
without dissipation until it is shocked at the jet-star discontinuity.
In fact, several recollimation shocks are evident in our simulations.
Such shocks reduce the jet speed and, as a consequence, also the speed
of the head (see Eq.~\ref{eq:bh}).

An alternative solution can be found by considering the ram pressure
of the inner part of the jet, which is in free expansion, against the
shear layer at the contact between the jet and star materials (Morsony
et al. 2007; Bromberg et al. 2011b). From the results in Table~1 of
Bromberg et al. (2011b) one can easily find:
\begin{equation}
\beta_h=\frac{L_j^{1/3}}{c\rho_a^{1/3}r_h^{2/3}\theta_0^{4/3}}
\label{eq:bromberg}
\end{equation}
The dependence of the jet head velocity on the jet luminosity is shown
in Figure~\ref{fig:f2} with thin black lines. The results of our
simulation are not able to distinguish between the derivation
presented here (Eq.~\ref{eq:final}) and the Bromberg et al. (2011b)
result. In either case, the jet head velocity is overestimated by a
factor $\sim4$ while the dependence on the jet luminosity is
qualitatively reproduced.

\section{Discussion and Conclusions}

We presented a set of simulations aimed at exploring the role of the
duration of the engine activity in the explosive phenomenology of
compact massive progenitor stars. Our simulations suggest that the
diversity in explosive outcomes (GRBs, weak-GRBs, X-ray flashes,
relativistic radio-bright SNe) is not necessarily reflected by an
analogous diversity in progenitor and engine properties. As a matter
of fact just one of our simulation sets, for a given progenitor
structure and a given total energy, can give examples of the
whole zoo of engine-driven explosions, as long as the engine activity
can have a wide range of durations, as observed in the GRB $T_{90}$
distribution (Kouveliotou et al. 1983).

\begin{figure}
\plotone{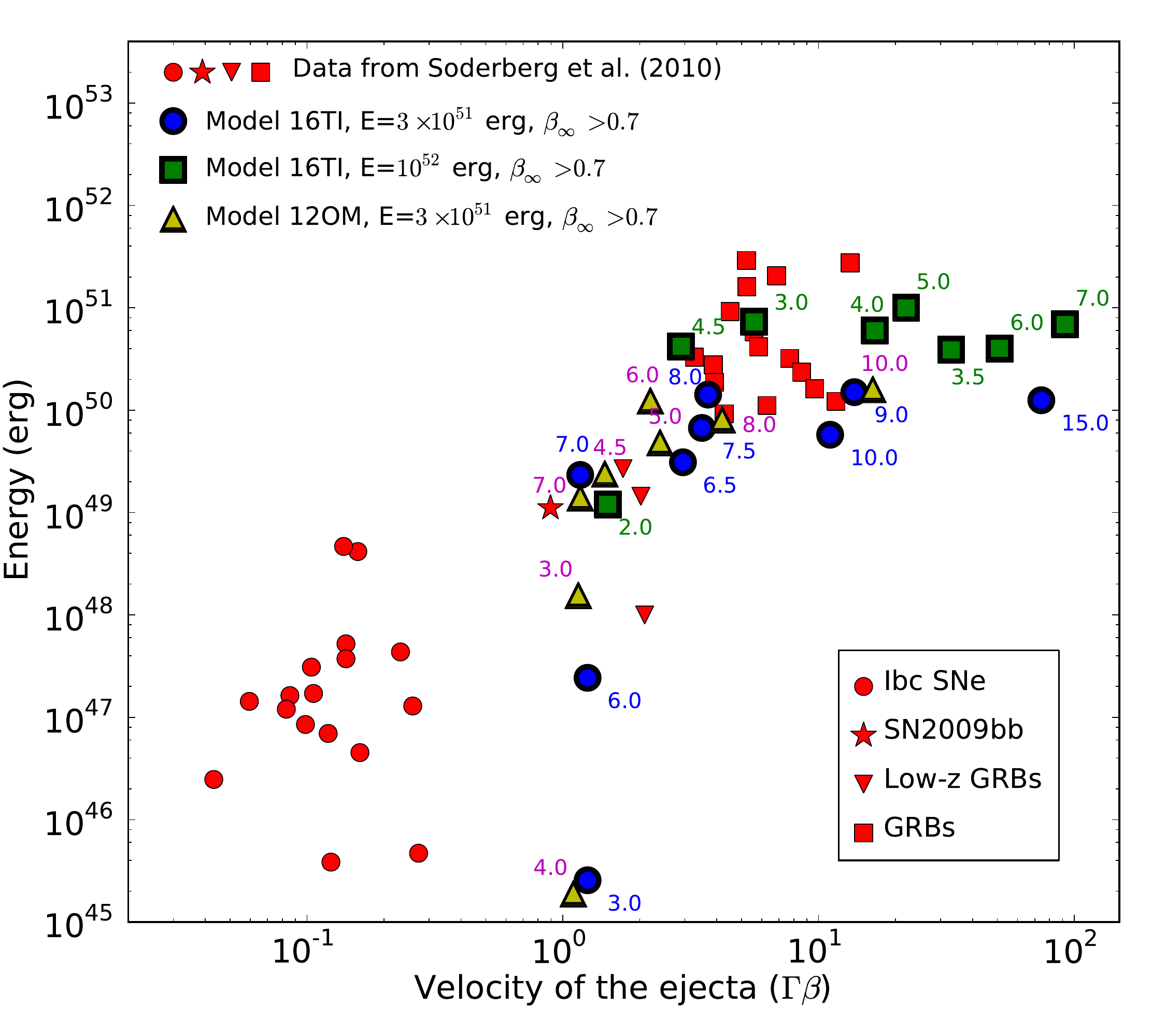}
\caption{{Comparison of the simulation results with the observational
    properties of Type Ibc SNe and jet-driven explosions (red
    symbols).  The results of our simulations are overlaid with blue,
    green, and yellow symbols (circles, squares, and triangles,
    respectively) with the duration of the engine activity of each
    simulation indicated next to each point in the corresponding
    color.}
\label{fig:f4}}
\end{figure}

\begin{figure}
\plotone{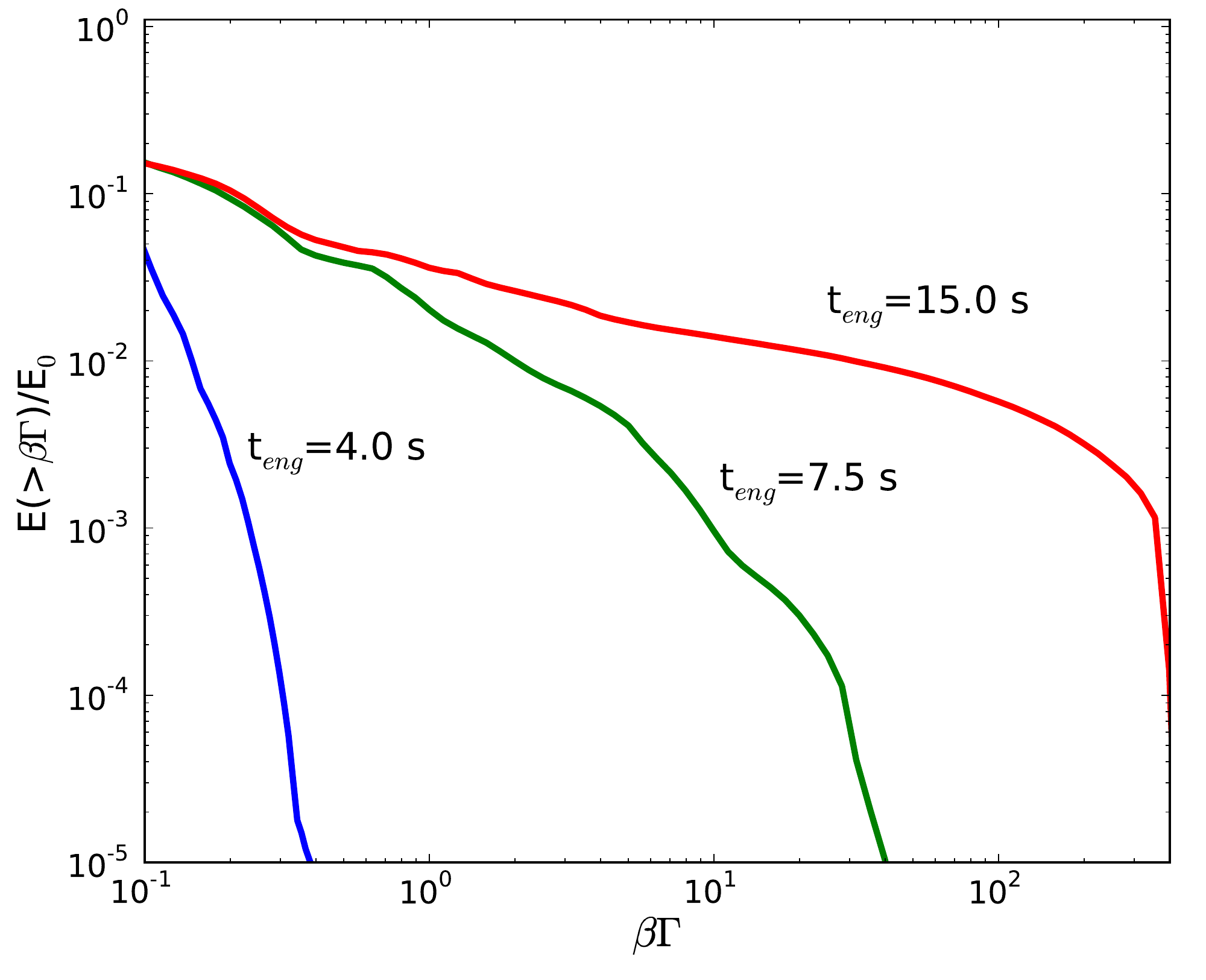}
\caption{{Cumulative energy distribution of three representative cases
  of relativistic jet-driven SN explosions. Simulations resulting in a
ordinary SN (blue line), a relativistic SN/weak GRB (green line) and a
full-fledged GRB (red line) are shown. All the three simulations have
progenitor 16TI and total engine energy $E=3\times10^{51}$~erg.}
\label{fig:edistri}}
\end{figure}

Jet-driven explosions like the ones we explore, however, do not seem
to be able to reproduce the properties of the ordinary type Ibc SNe in
the lower left corner of Figure~\ref{fig:f4}.  This is expected, since
stellar evolution models predict that only a small fraction of massive
star have the properties required to harbor a GRB engine in their
center (Woosley \& Heger 2006; see, however, Papish \& Soker 2001 for
an alternative opinion).  However our simulations do predict that if
jet engines with short active times exist they produce stellar
explosions.  For example the $t_{\rm{eng}}=3.0$, 4.0, and 5.0 s
simulations for the 16 solar-mass progenitor and total energy
$E=3\times10^{51}$~erg resulted in 90\% of the stellar material
unbound.  Such explosions would not be associated with fast, bright
radio transients.  A member of this new class of explosions may have
been recently identified in SN2010jp (Smith et al. 2011).  It must be
noted, however, that the comparison of our simulations with
observations in the lower left corner of Figure~\ref{fig:f4} is
severely affected by the somewhat arbitrary choice of a lower limit
$\beta\ge0.7$ for the material to be included in the calculation of
the ``relativistic'' ejecta. As shown in Figure~\ref{fig:edistri}
(blue line), the energy distribution of ordinary-looking jet-driven
SNe is very steep and a change in the velocity threshold affects
dramatically the amount of energy labeled as ``relativistic''. In
addition, the data from the simulations are extracted just a few tens
of seconds after the engine onset, while the radio data used by
Soderberg et al. (2010) were taken on a time scale of weeks after the
burst trigger. In order to accurately compare the weakest radio
transients in Figure~\ref{fig:f4} to our simulations, it is required
to evolve the simulations much farther in time and space to directly
compute the radio emission, a task that is not feasible at the
required resolution with current instrumentation.  Additional
signatures, such as asphericity, nucleosynthesis patterns, absorption
and emission lines profiles, and linear polarization need to be
explored in order to confirm our results and find observable
quantities that could allow us to distinguish between radio-quiet
jet-driven explosions and core-collapse SNe.

Our simulations predict that engines with a long duration always
produce a successful burst, while only within a certain duration
interval they produce events of the intermediate classes, such as
relativistic Ibc SNe and weak GRBs or X-ray flashes. Our 16TI
progenitor, for example, produces intermediate events for $6\le
t_{\rm{eng}}\le10$~s and $2\le t_{\rm{eng}}\le4.5$~s for the
low-energy and high-energy engines, respectively. If the distribution
of engine durations were flat, our mode would predict a larger rate of
successful GRB events compared to events of the intermediate
class. Such prediction would be inconsistent to current estimates,
estimating the current rate of intermediate events either comparable
to that of successful events (Soderberg et al. 2010) or an order of
magnitude larger (Soderberg et al. 2006). Such comparison led Bromberg
et al. (2011a) to conclude that low-luminosity GRBs have a different
origin from successful GRBs. However, the intrinsic engine duration
distribution is not known, and it may well be that due to the high
angular momentum that is required in the collapsar model to sustain
prolonged accretion onto the central compact object (MacFadyen \&
Woosley 1999, Woosley \& Heger 2006), the intrinsic distribution is
skewed towards short durations, explaining the high rate of
intermediate class events. Further observations and a better
theoretical understanding of the powering of relativistic jets will
help make this constraint more useful for future comparison. It
should also be pointed out that the viewing angle does have an
influence on the burst energetics and spectrum for a given
progenitor-engine pair (e.g., Lazzati et al. 2011), adding to the
diversity of the events.

Our simulations also predict a correlation between the engine duration
and the energy observed as electromagnetic radiation. When trying to
compare this prediction with observations, one should keep in mind
that the engine duration is not equal to the duration of the prompt
emission (the $T_{90}$) and that the energy in radiation is the ``true
energy'', not the isotropic equivalent one. It is particularly
difficult to connect the engine duration to the $T_{90}$ duration. In
most cases of successful bursts one has $t_{\rm{eng}}>T_{90}$, but the
difference between the two quantities can be large. For example,
Lazzati et al. (2010) find that the same engine observed from
different viewing angles can give rise to GRBs with durations spanning
from a tenth of a second up to 100 seconds (the duration of the engine
in that particular simulation). Such a diversity of durations of the
radiative phase for a given engine duration is due to the interaction
of the jet with the progenitor star. Observers very close to the jet
axis see an initially very bright phase, due to the hydrodynamic
collimation of the energy along the jet axis. Observers at larger
angles, however, do not see any emission at early times since the jet
has been collimated to an opening angle smaller than the angle of the
line of sight. On the other side of the duration distribution, short
engines can produce relatively long bursts or flashes since the
dynamics of the ejecta is very different between events dominated by
the shock breakout (e.g. Campana et al. 2006; Nakar \& Sari 2012) and
events dominated by fully developed, highly relativistic outflows.

\begin{figure}
\plotone{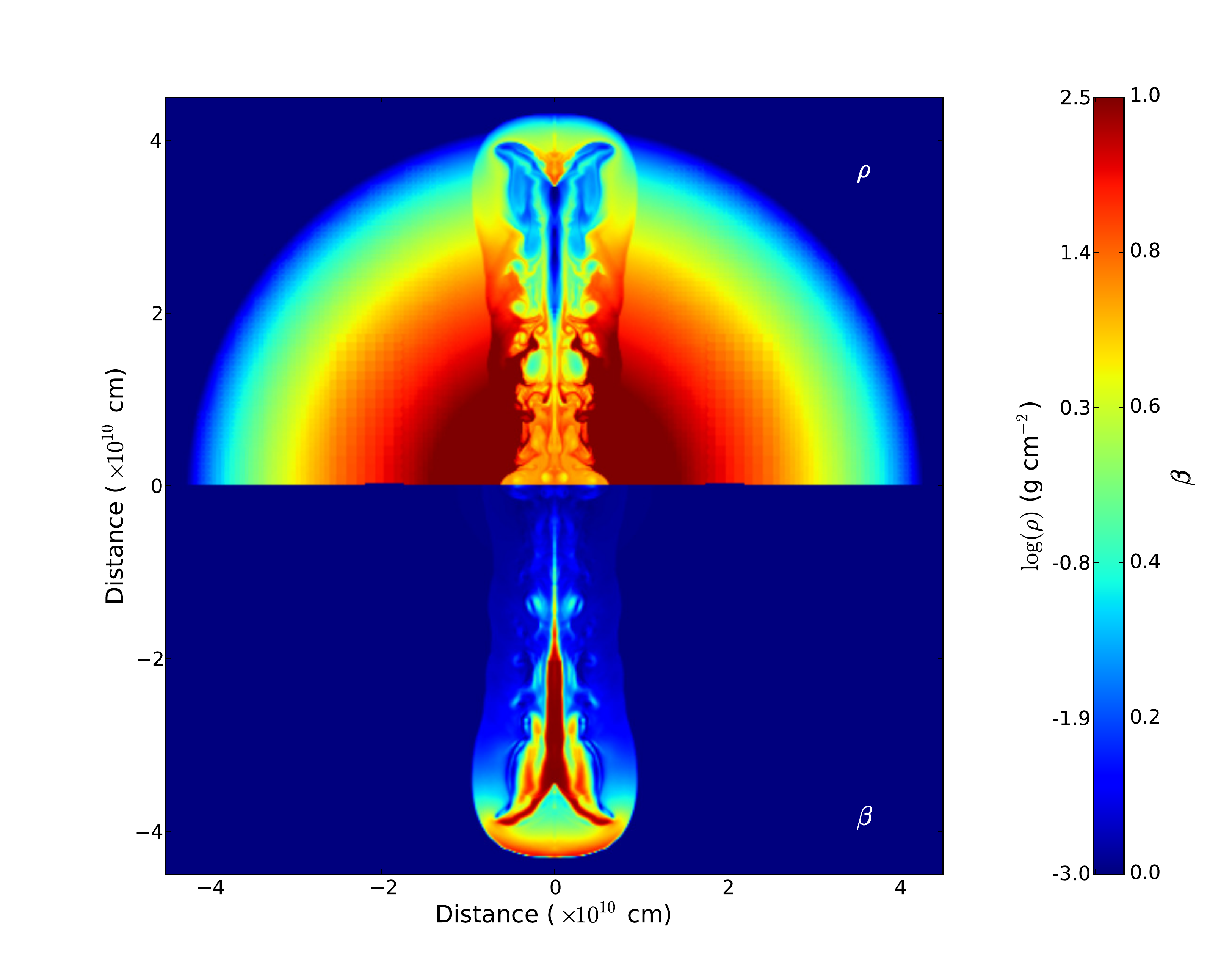}
\caption{{Density and velocity maps for the $t_{\rm{eng}}=7.5$~s
    simulation at breakout ($t=8.13$~s).  The top panel shows a
    false-color rendering of the logarithm of the density, while the
    bottom panel shows velocity in units of the speed of light (see
    color scales on the right).}
\label{fig:f1}}
\end{figure}

As a final remark, it should be noted that the simulations that we
have presented here do not include all the complex physics of a
stellar explosion. The most important limitations are likely the facts
that magnetic fields are not considered, that the jet is injected at a
somewhat large radius, that the simulations are performed in 2D, and
that the engine/jet properties are not self-consistently derived from
the progenitor properties. All these limitations are likely to affect
the details of the jet propagation. For example, a well-known 2D
instability produces the wedge of high-density, slow-moving material
that is visible ahead of the jet in Figure~\ref{fig:f1}. Such a wedge
disappears in 3D simulations (Zhang et al. 2004) and its appearance in
2D simulations has the effect of making the jet propagation slightly
slower. The bottom line is that the precise numerical values of the
engine durations that correspond to each class of explosions are not
to be taken at face value. Future, more refined simulations will
likely update those values and are required to further evaluate the
role of the engine duration in explaining the zoo of jet-driven
stellar explosions.

\acknowledgements Resources supporting this work were provided by the
NASA HEC program through NAS. The software used in this work was in
part developed by the DOE-supported ASC / Alliance Center for
Astrophysical Thermonuclear Flashes at the University of Chicago. This
work was supported in part by NASA Astrophysics Theory Program grant
NNX09AG02G (MCB), NSF grant AST-0907872 (MCB), and Fermi GI program
grant NNX10AP55G (DL \& CHB). BJM is supported by an NSF Astronomy and
Astrophysics Postdoctoral Fellowship under award AST-1102796.

\newpage

\newpage

\newpage

\newpage

\newpage

\newpage

\newpage

\end{document}